# Assertion-Based Design Exploration of DVS in Network Processor Architectures

Jia Yu, Wei Wu, Xi Chen, Harry Hsieh, Jun Yang
University of California, Riverside
{jiayu, wwu, xichen, harry, junyang}@cs.ucr.edu

Felice Balarin
Cadence Berkeley Laboratories
felice@cadence.com

## Abstract

*With the scaling of technology and higher requirements on performance and functionality, power dissipation is becoming one of the major design considerations in the development of network processors. In this paper, we use an assertion-based methodology for system-level power/performance analysis to study two dynamic voltage scaling (DVS) techniques, traffic-based DVS and execution-based DVS, in a network processor model. Using the automatically generated distribution analyzers, we analyze the power and performance distributions and study their trade-offs for the two DVS policies with different parameter settings such as threshold values and window sizes. We discuss the optimal configurations of the two DVS policies under different design requirements. By a set of experiments, we show that the assertion-based trace analysis methodology is an efficient tool that can help a designer easily compare and study optimal architectural configurations in a large design space.*

## 1 Introduction and Motivation

As Internet gets more and more complicated with the rise of new protocols and services, so does the cost of new equipment and upgrades. A network processor (NPU) is a base hardware platform that provides high performance and flexible programming capabilities, which allows it to address many market segments and a wide range of applications. As a result, the cost of upgrade can be reduced and developing cycles for new protocols and data types can be shortened. Therefore, NPUs are poised to replace expensive and inflexible fixed-function silicon application-specific integrated circuits (ASICs).

A number of challenges for NPU implementation are already evident, and power dissipation is among one of them. For example, in a typical router configuration, there may be one or two NPUs per line card. A group of line cards, e.g. 16 or 32, are generally placed within a single rack or cabinet. Thus, the aggregated heat dissipation becomes a big concern, given that each NPU typically consumes around 20 Watts and the operating temperature can reach as high as $70^oC$ [13]. On the other hand, with the demand of performance scaling, NPU's clock frequency is increasing and more computation engines will be put on an NPU. Figure 1 shows the power and performance changes in three Intel IXP family NPUs [10, 12, 13]. Note that the power dissipation increases as the complexity of NPU increases. This trend brings significant challenges for the NPU design.

| Description | IXP1200 | IXP2400 | IXP2800 |
|---|---|---|---|
| Performance(MIPS) | 1200 | 4800 | 23000 |
| Media Bandwidth(Gbps) | 1 | 2.4 | 10 |
| Frequency of ME(MHz) | 232 | 600 | 1400 |
| Number of MEs | 6 | 8 | 16 |
| Power(W) | 4.5 | 10 | 14 |

Figure 1: The power and performance of Intel IXP NPUs.

System level modeling with executable languages such as C/C++ or other modeling frameworks have been crucial in designing large electronic systems. Unfortunately, most cycle-level accurate simulators only report performance and power data for worst and/or average cases, which pose limitation on power/performance analysis. For example, an NPU's performance and power dissipation are closely related to the workload, namely the incoming packet rate. The workload is usually unbalanced, which may cause extreme high power dissipation occasionally. The unbalanced workload provides opportunities for power and performance tuning. The power and performance distribution patterns are important complements to average/worst-case data in the design exploration.

It has been shown that the assertion-based analysis methodology is very suitable for transaction-level or cycle-level design exploration, specially in power/performance analysis of NPU designs. The basic methodology has been proposed in [6] for verifying and analyzing basic functional and performance properties of an NPU design. From formally specified assertions, trace checkers and distribution analyzers are automatically generated to validate or analyze simulation traces. Designers do not need to write separate reference models or scripts to scan through the traces. So it is very suitable for design exploration of large systems with high complexity and functionality such as NPU designs.

In this paper, we focus on the assertion-based design exploration of dynamic voltage scaling techniques in the NPU model. In order to efficiently analyze the power-performance trade-offs among different DVS policies with different parameter settings, we use Logic of Constraints (LOC) [4] to specify assertion formulas for power and performance distributions.

With automatically generated distribution analyzers, we compare their power and performance characteristics and identify optimal configurations in their large design spaces.

The rest of the paper is organized as follows. In the next section, we introduce the network processor model, the basic DVS technique and the assertion-based trace analysis methodology. In Section 3, we describe the experiment settings for the network processor simulator NePSim including benchmarks, IP traffic files,and simulation traces. In Section 4, we present the procedures and analysis results of assertion-based design exploration for DVS policies in the NPU model. We compare two types of DVS techniques, traffic-based DVS and execution-based DVS, with different parameter settings, and find optimal configurations for both DVS techniques under different design requirements. Section 5 concludes the paper.

# 2 Background

## 2.1 Network Processor Model

A network processor design usually contains multiple RISC processing cores, dedicated hardware for common networking operations, high-speed memory interfaces, high-speed I/O interfaces, and interfaces to general purpose processors. Here we use NePSim simulator [8] to model the NPU architecture. NePSim is based on Intel IXP1200 and includes a cycle-accurate architecture simulator and a power estimator. All the configurations in NePSim are parameterizable.

The reference model of the network processor design follows IXP1200 and consists of a StrongARM core, six multi-threaded processing units called microengines (MEs), memory controllers, high-speed buses, and packet buffers. The StrongARM core initializes the microcode program to control stores of the microengines and loads necessary data into memory before enabling the microengines. The off-chip SRAM (up to 8M) is typically used to store the forwarding table, while the SDRAM (up to 256M) is typically used to store IP packets. The usage of each component is highly dependent on the application and workload.

## 2.2 Dynamic Voltage Scaling

Dynamic voltage scaling (DVS) [3] is a popular low power technique and has been employed widely for microprocessors, resulting in significant power and energy savings. DVS exploits the variance of a processor's utilization, reducing voltage and frequency (VF in short) when the processor has low activity and increasing VF when the peak performance is required. Dynamic power consumption is proportional to $C \cdot Vdd^2 \cdot \alpha \cdot f$, so reducing voltage ($Vdd$) and frequency ($f$) can significantly reduce power consumption.

Although many DVS algorithms appear in literature, the unsolved difficulty is how to derive the optimal settings from external observations, for example, by monitoring the workload or idle time. In this paper, we will use assertion-based methodology to study and find out optimal DVS parameters in NPUs.

## 2.3 Assertion-Based Analysis Methodology

Assertion-based checking is similar to the popular embedded assertion technique in hardware design, where simple comparison circuitry is inserted into HDL descriptions to help designers uncover bugs during simulation. The methodology begins with a formula, e.g. in Logic of Constraints (LOC), and generates stand-alone checkers, independent of any simulation language and platform [4, 6]. Furthermore, LOC is designed to specify quantitative performance and functional properties for analysis of transaction-level execution traces. The basic components of LOC are *event names*, *instances of events*, *annotations*, and a single *index variable i*. For example, a latency property (a dequeue event happens no later than 50 cycles after the corresponding enqueue) can be formally specified as an LOC formula: *cycle(deq[i])- cycle(enq[i])<=50*. The formula is satisfied if it holds for all event instances, i.e. for all values of $i$. The automatically generated checkers are used to analyze simulation trace files and report all the violations of the assertions.

To automate quantitative distribution analysis that is common in design exploration, we extend the LOC assertions by introducing 3 more operators $\bowtie$, $\lhd$ and $\rhd$. To analyze the distribution of some quantity over certain ranges, we can use a formula, in the form of $quantity \bowtie \{min, max, step\}$, to automatically generate a corresponding analyzer. An *analysis period* is specified with a triple $\{min, max, step\}$, where *min* and *max* are lower and upper bounds, and the interval between these two values is divided into bins of width *step*. For example, given a formula:

$$(time(forward[i+100]) - time(forward[i])) \bowtie \{40, 80, 5\} \quad , \qquad (1)$$

an assertion analyzer is generated to evaluate the left hand side with $i$ being 0, 1, 2, ... , and report the percentage of formula instances whose values fall within the ranges of $(-\infty, 40]$, $(40, 45]$, ..., $(75, 80]$, $(80, +\infty)$. If we replace the operator $\bowtie$ with $\lhd$ or $\rhd$, the ranges become $(-\infty, 40]$, $(-\infty, 45]$, ..., $(-\infty, 75]$, $(-\infty, 80]$ or $[40, +\infty)$, $[45, +\infty)$, ..., $[75, +\infty)$, $[80, +\infty)$, respectively.

# 3 Experimental Settings

In this section, we introduce our experiment environment, IP packet traffic models used in the simulation, and the simulation traces.

## 3.1 Benchmark Applications

In our experiments, we choose four representative networking applications to explore different architectural features of the NPU model, i.e. *ipfwdr, url, nat* and *md4*. The application *ipfwdr* is an IP forwarding software provided in Intel's

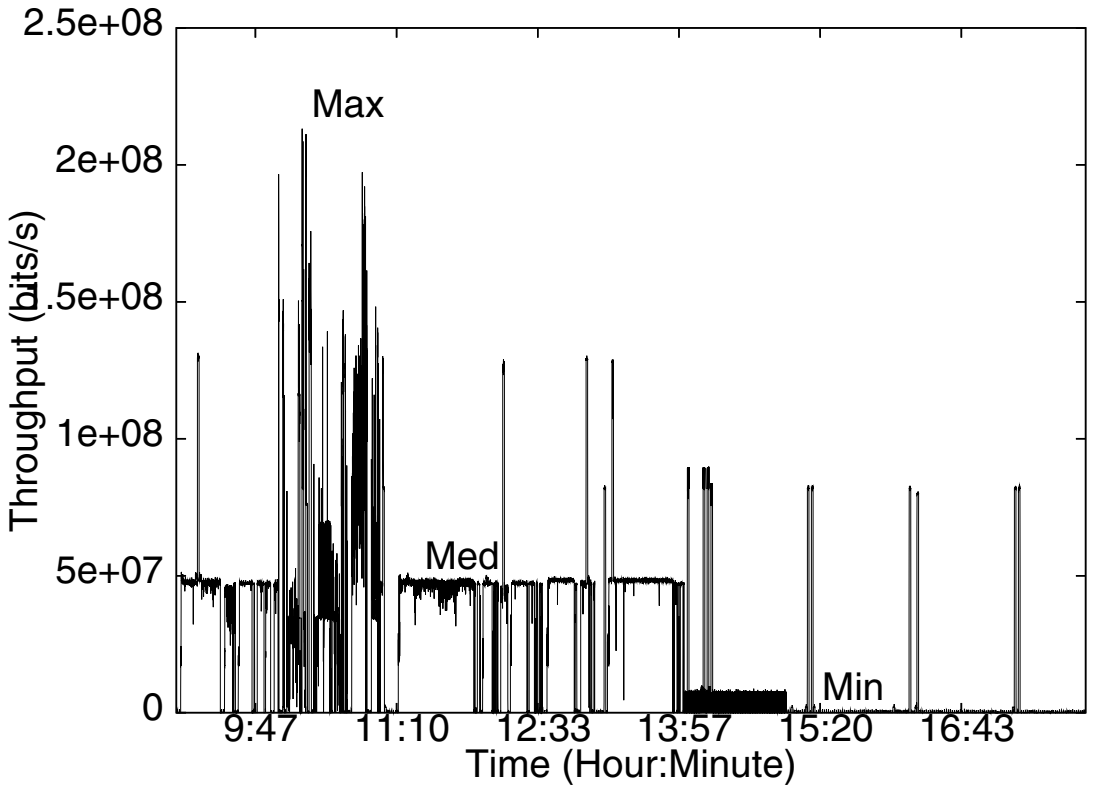

Figure 2: Example IP packets distribution

| Event type | Details |
|---|---|
| *pipeline* | an instruction enters the execution pipeline |
| *forward* | an IP packet is forwarded |
| *fifo* | an IP packet is put into the processing queue |
| **Annotation type** | **Details** |
| *cycle* | number of core clock cycles elapsed from the beginning |
| *time* | simulated time elapsed from the beginning |
| *energy* | cumulative energy consumed |
| *total_pkt* | total packets received or transmitted |
| *total_bit* | total bits received or transmitted |

Figure 3: List of event and annotation types.

| cycle | time(us) | energy | p_loss | event |
|---|---|---|---|---|
| | | ... ... | | |
| 365 | 1.573 | 0.768133 | 120 | m2_pipeline |
| 366 | 1.577 | 0.773932 | 120 | m3_pipeline |
| 367 | 1.580 | 0.784506 | 121 | forward |
| 368 | 1.583 | 0.794108 | 121 | m5_pipeline |
| 369 | 1.587 | 0.809369 | 121 | m4_pipeline |
| | | ... ... | | |

Figure 4: A snapshot of NePSim simulation trace.

SDK. The routing table is stored in the SRAM and the output port information is stored in the SDRAM. The program *url* routes packets based on URL requests. It checks the payload of packets frequently, so it needs a large number of SRAM and SDRAM accesses. In *nat* (network address translation), each packet only needs an access to SRAM for looking up the IP forwarding table. The *md4* provides a 128-bit digital signature algorithm. It moves data packets from SDRAM to SRAM and accesses SRAM multiple times for computation. It is therefore both memory and computation intensive.

Memory accesses, specially SDRAM accesses, have long latency. They lead to long idle time for MEs, which in turn shows up as lower power and throughput. Computation-intensive benchmarks, those that do not wait on memories, will tend to show higher power consumption.

### 3.2 IP Traffic Patterns

The simulation inputs follow IP packet traffic patterns in a real world edge router from NLANR [15]. Figure 2 shows a day time distribution of IP packet arriving rates. It is obviously too expensive to simulate the entire day's worth of simulation traces for the purpose of design space exploration. We sample a few seconds of real traffic in high, medium and low arriving rates as individual inputs to the simulator.

### 3.3 Simulation Traces

The simulator provides the assertion analyzer with necessary data traces. The traces contain a set of architectural execution events that occur frequently during simulation and a set of power/performance related values called annotations. In our experiments, we mainly use three types of events, *pipeline*, *forward* and *fifo*, explained in Figure 3. In a simulation trace, the events are prefixed to differentiate different microengines (MEs) or configurations. For example, *m2_pipeline* represents a pipeline event from ME2. Each event is associated with five annotations (see Figure 3). A snapshot of a trace file generated by NePSim simulator is shown in Figure 4.

## 4 Design Exploration of DVS

In a real system with DVS, the frequency and voltage are adjusted dynamically according to the processing workload. A DVS scheduler relies on the history information of workload to make decisions. In an NPU design, two types of information can be used for this purpose, network traffic load and processor idle time. We call the two DVS policies traffic based dynamic voltage scaling (TDVS) and execution based dynamic voltage scaling (EDVS). We do not combine the two policies because monitoring both traffic load and processor idle time on a chip is expensive in terms of area and power.

In this section, we analyze the power/performance trade-offs of DVS policies by varying the window size and threshold for voltage/frequency scaling, and search for optimal points in the design space. We also compare the two DVS policies through their power and performance results under different design requirements.

### 4.1 Traffic based Dynamic Voltage Scaling

TDVS uses the total traffic load detected at the 16 device ports as the control parameter for scaling. If the traffic volume in the previous time window is smaller or larger than a particular threshold value, we scale down or up the VF of the processor by one step, until a lower or upper bound is hit. The lower and upper bounds of VF, similar to those used in Intel XScale [11], are from 400MHz to 600MHz and 1.1V to 1.3V. We set the frequency step to 50Mhz and compute the voltage as in XScale. In order to match higher NPU frequency, we scale the speed of SDRAM, SRAM and ixbus to 1.3 times of those in IXP1200.

To estimate the power in TDVS, we modified NePSim's power estimation module to include the power overhead, a 32-bit adder. The adder is used to accumulate the packet sizes in each monitor window, and compare the traffic volume with the threshold. Note this adder is only used when a packet comes in, much less frequently than the ALUs in ME pipelines. From the experiment results, we find the overhead is less than 1% of total power.

TDVS reduces the power, but it may adversely affect the performance. The clock cycle becomes longer if *Vdd* is decreased, so the NPU takes longer time and possibly more energy to get the same amount of work done. The trade-off motivates us to analyze both power consumption and performance of the NPU with different TDVS policies applied. The goal is to find the optimal points in the design space for each benchmark. We use the following LOC formula to analyze the power consumption distribution:

$$(energy(forward[i+100]) - energy(forward[i]))/ \\ (time(forward[i+100]) - time(forward[i])) \\ \triangleright \{0.5, 2.25, 0.01\} \ . \quad (2)$$

The left hand side of the formula calculates the average power consumption for each 100 packets forwarded.

To study the performance of the processor with various configurations, we analyze the distribution of the transmitting throughputs using the following formula:

$$((total_bit(forward[i+100]) - total_bit(forward[i]))/10^6) \\ /(time(forward[i+100]) - time(forward[i])) \\ \triangleleft \{100, 3300, 10\} \ . \quad (3)$$

The left hand side of the formula calculates the average forwarding bit rate in Mbps for each 100 packets forwarded.

With the two formulas, we search for the optimal settings of TDVS policies. In TDVS, two main types of parameters that need to be carefully tuned are the traffic thresholds and window size. For each TDVS policy, the traffic thresholds are a set of volume numbers that control the voltage scaling in different VF combinations. With the frequency and voltage reduced, the traffic threshold is also lowered to match the reduced ME processing capability. Taking *ipfwdr* as an example, we choose a top threshold of 1000Mbps for the normal frequency of 600MHz and other thresholds for reduced VFs are decided as shown in Figure 5. In our experiments, we use the benchmark *ipfwdr* to compare the TDVS policies with four different top thresholds: 800, 1000, 1200, and 1400 Mbps.

| Frequency (Mhz) | 600 | 550 | 500 | 450 | 400 |
|---|---|---|---|---|---|
| Voltage(V) | 1.3 | 1.25 | 1.2 | 1.15 | 1.1 |
| Traffic Threshold(Mbps) | 1000 | 916 | 833 | 750 | 666 |

Figure 5: The detailed scaling values.

The window size decides how long a traffic history is used to make voltage scaling decisions, and it also directly affects the overall performance of the TDVS policy. For example, if the window size is set to 20k clock cycles, the average traffic volume in the previous 20k cycles is compared to the current threshold to decide whether the VF needs to be changed. If a window size is too large, it may smooth the peak traffic with low traffic and miss a good chance to reduce power; If window size is too small, VF may change too frequently, which incurs more penalty and eventually hurts the performance. In our experiments, the penalty for each voltage scaling is 10us [8], which is equivalent to 6000 cycles at the normal frequency of 600MHz. We compare 4 different window sizes for *ipfwdr*, ranging from 20k to 80k cycles.

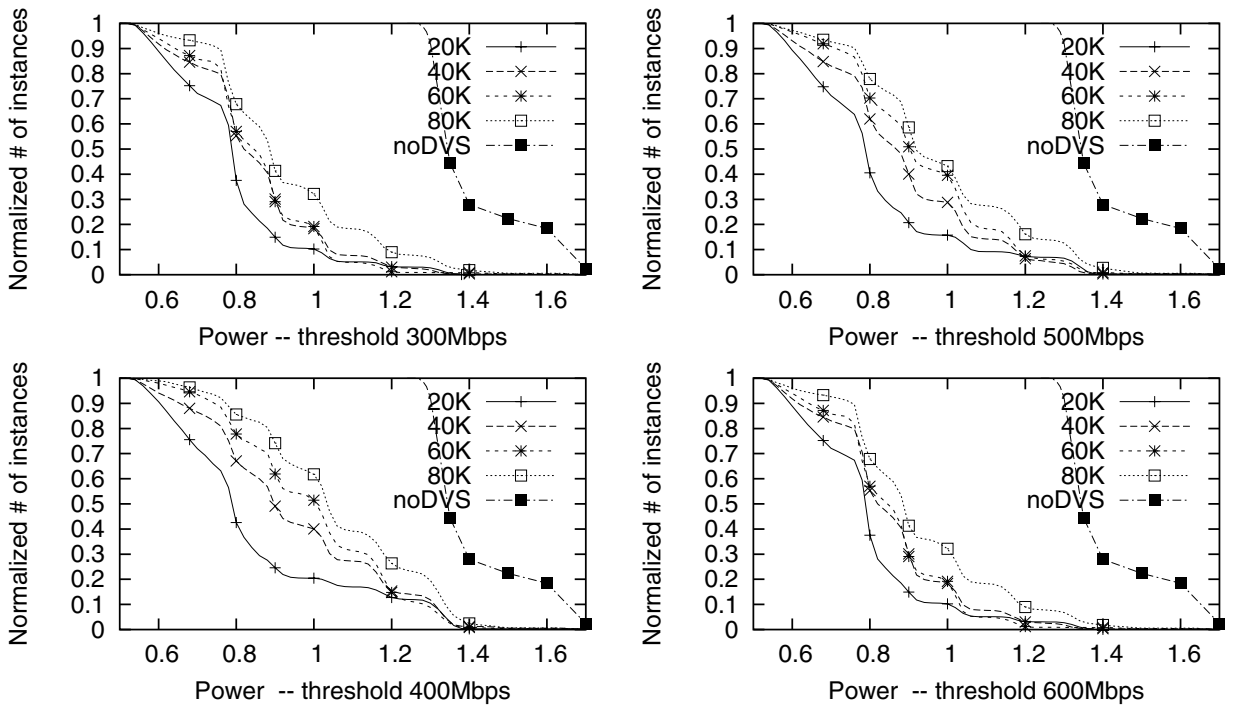

Figure 6: Power under different design points with TDVS.

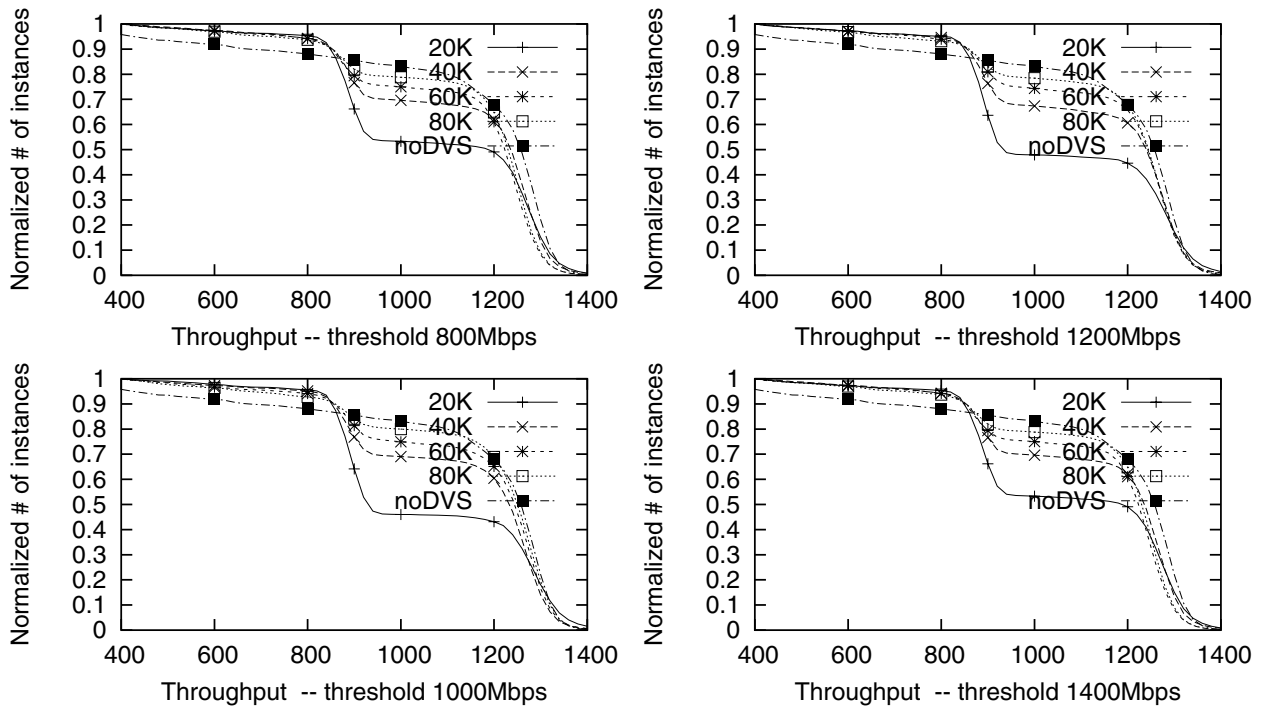

Figure 7: Throughput under different design point with TDVS.

We run the simulation $8 \times 10^6$ cycles for each TDVS configuration. Using the automatically generated distribution analyzer with the formulas (2) and (3), we compare the power and performance distributions with different TDVS policies or no TDVS enabled. The distributions for the power and performance are plotted in Figure 6 and Figure 7 respectively. Each subgraph shows the power or throughput distribution with a particular top threshold and different window sizes. In the power distribution graphs, the horizontal axis represents possible power values and the vertical axis represents the percentages of assertion instances that are smaller than particular power values. Similarly, in the throughput distributions, the vertical axis represents the percentages of assertion instances that are larger than particular throughput values.

From Figure 6, we can see that compared with no TDVS policy, the power saving by TDVS is obvious no matter what threshold or window size is chosen. In most cases (except with window size of 20k), the performance degradation is small (from Figure 7). It is therefore shown that TDVS is a very successful power saving technique. We also see that TDVS configurations with smaller window sizes have lower power consumption but worse throughput, regardless the threshold values. When window size is small, e.g 20k, the TDVS policy becomes very aggressive. The VFs are changed very frequently, and as a result, the 6000-cycle penalties almost consume 30% of the window time. That is the reason why there is dramatic drop in throughput for window sizes of 20k. On the other hand, for 80k window sizes, certain power savings are still achieved with almost no performance loss.

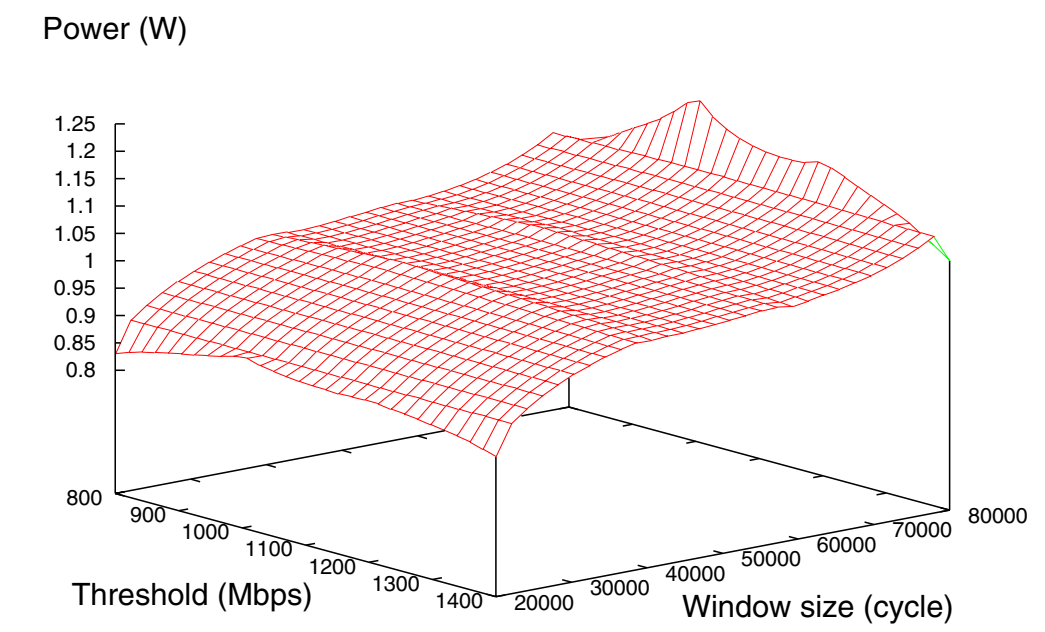

Figure 8: Power under different design points with TDVS.

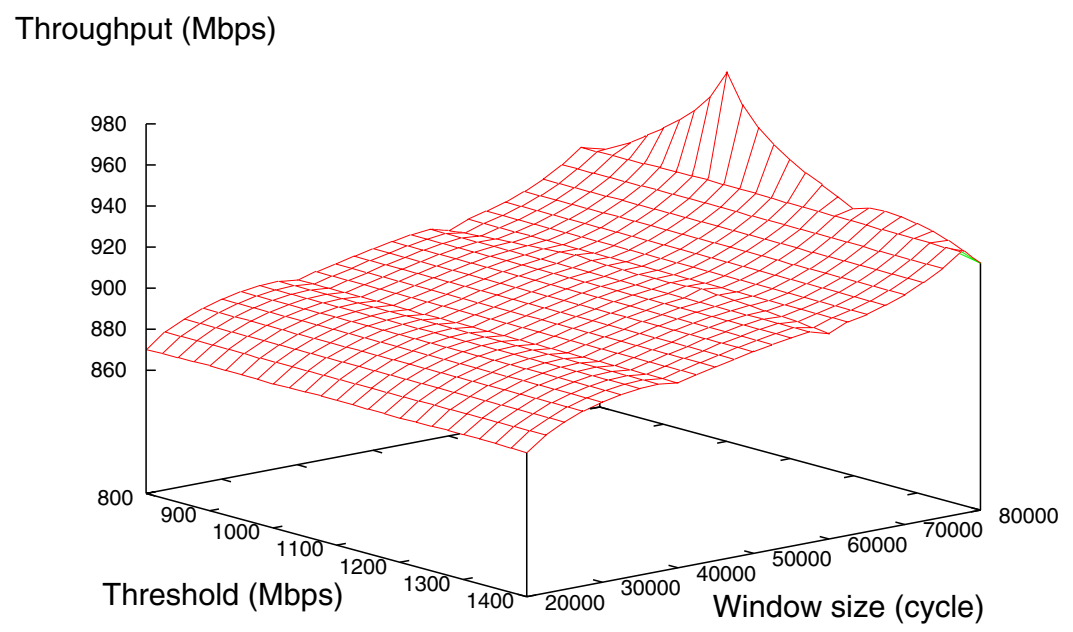

Figure 9: Throughput under different design point with TDVS

To compare the results of different thresholds more clearly and look for a best TDVS policy for *ipfwdr* with an optimal threshold-window size combination, we generate 3-D graphs for power and performance distributions in Figure 8 and Figure 9. A vertex on the surface shown in Figure 8 represents that 80% of formula (2) instances are lower than a power value for a particular threshold and window size. Similarly, a vertex on the surface in Figure 9 represents that 80% of formula (3) instances are higher than a throughput value for a particular threshold and window size. As shown in Figure 8, for a particular window size, the threshold of 1000Mbps has higher power than others, and this trend becomes more significant as the window size increases. As shown in Figure 9, if the window size is small, the performances for different thresholds are similar; as the window size becomes larger, the performance for 1000Mbps threshold becomes much better than others.

Based on above analysis, if performance has a higher priority in the design, we should choose threshold of 1000Mbps and 80k window size resulting in limited power savings. On the other hand, if saving power is more important, the configuration with 1400Mbps and 40k of window size is preferred. And this result is specific to this particular *ipfwdr* application.

## 4.2 Execution based Dynamic Voltage Scaling

In execution based dynamic voltage scaling (EDVS), the idle time of microengine is used as the control parameter for voltage scaling. When the idle time is longer or shorter than a certain percentage of an observed period, the VF of the microengine is scaled down or up by one step, until a lower or upper bound is hit. Note that in EDVS, each ME changes its VF independent. Intuitively, ME idle time is usually seen to be proportional to the workload, which makes TDVS and EDVS almost the same. However, this is not really the case in the NPU model. Even if an ME does not process packets during low workload, it will actively execute instructions to poll the buffers and status registers to check new packets. In the NPU model, the idle time of an ME is mainly introduced by long latency of memory accesses since an SDRAM access can take as much as 100 clock cycles. If all the threads in an ME are waiting for memory accesses to be completed, we consider the ME idle.

To analyze EDVS policies, the idle time thresholds and window sizes are the main parameters. Other parameters are configured as those used in TDVS. We use the assertion-based distribution analyzer to find the good idle time thresholds by analyzing the distribution of the idle time in simulations. It is observed that for receiving MEs, in around 90% of the total simulation time, idle time is either under 5%, or between 30% and 40%, indicating two modes of operation. For transmitting MEs, idle time is almost always under 5%, indicating a transimmion constrained scenario. The microengines seem working under only two statuses, either busy or idle. Here we simply choose the idle time threshold value as 10%, i.e. if the idle time of an ME is longer or shorter than 10% of an observed period determined by the window size, its VF may be changed. We study three different window sizes, 20k, 40k and 60k and still use *ipfwdr* as the example benchmark.

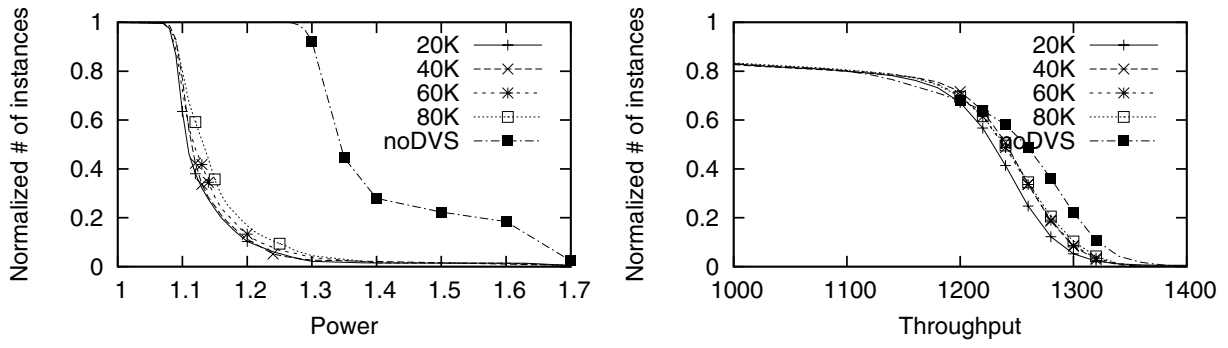

Figure 10: Power and performance distribution for EDVS

We run the simulation $8 \times 10^6$ cycles for each EDVS configuration and plot the distributions of throughput and power in Figure 10. From the power distribution graph, we observe that power dissipation generally drops from 1.5W to 1.15W for most cases with EDVS applied, achieving around 23% of power saving. Meanwhile, there is nearly no performance degradation from the throughput distributions. In EDVS, each ME changes its VF independently and the transmitting MEs never scales down their VFs due to their low idle time.

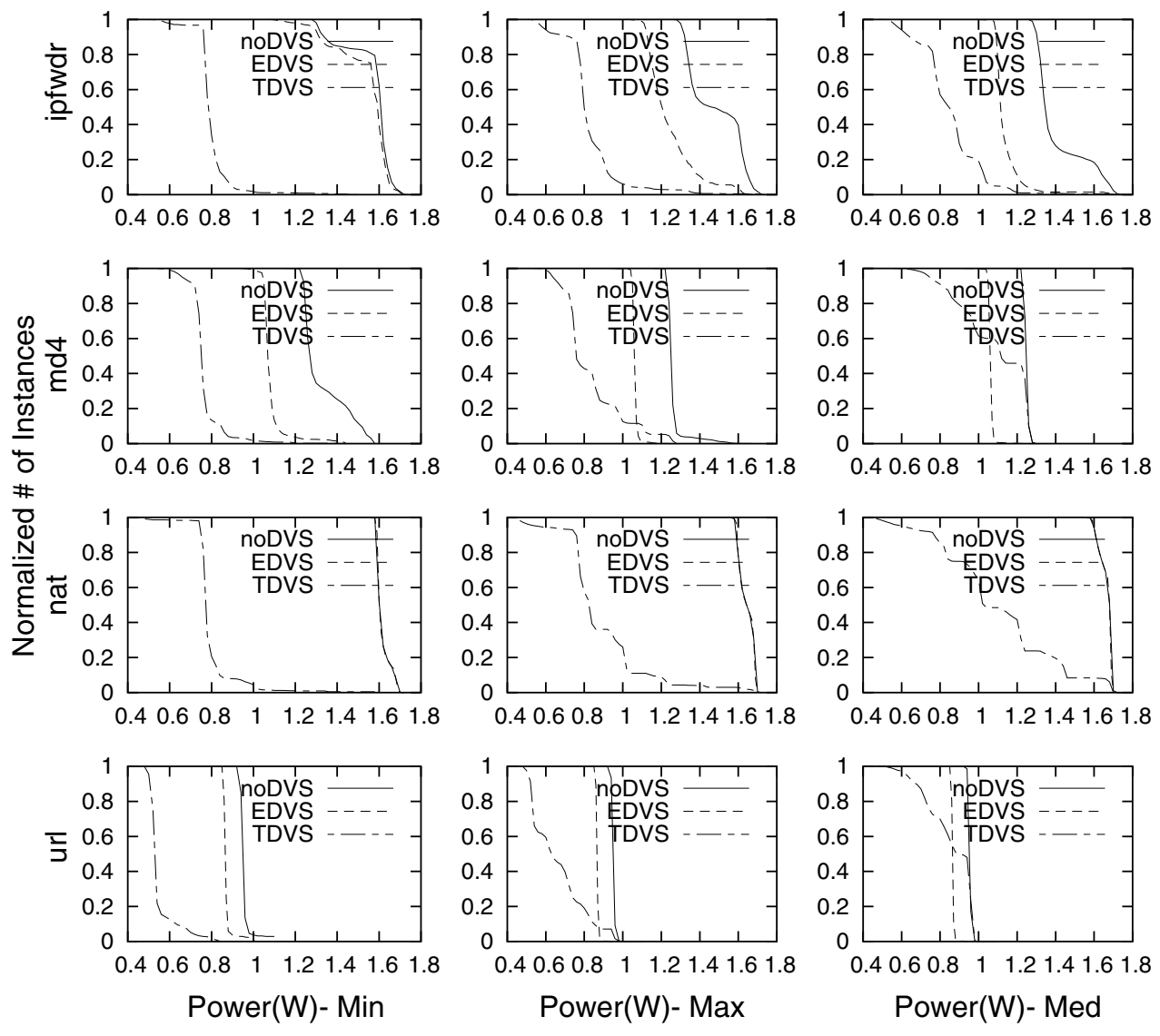

Figure 11: Energy comparisons for employing DVS

### 4.3 Comparison between TDVS and EDVS

We have shown that both TDVS and EDVS are capable of saving power with little performance sacrificed. Now we are ready to compare the two policies, and find which one is better given a particular power or performance requirement. We sample the real traffic file in three periods with high, medium, and low traffic volumes respectively. We simulate all four benchmarks with the optimal configurations (from previous analysis) for two DVS policies and compare the power distributions in Figure 11. We do not show the throughput performances and only note that in all cases EDVS has no significant performance loss while TDVS never drops more than 2-5% compared to the original NPU model with no DVS applied.

Overall, TDVS has more power savings than EDVS. But as the traffic volume becomes higher, power savings by TDVS reduce quickly, while EDVS has a more steady reduction under every situation. EDVS has better results for memory intensive benchmarks. We observe that *ipfwdr* shows the most power savings if traffic volume is medium or high. This is because *ipfwdr* needs to check routing tables in SRAM and the output port information in SDRAM for each packet. There are plenty of opportunities for EDVS. The benchmark *nat* shows no power savings from EDVS under every traffic patterns due to the fact that *nat* has very few memory accesses, and the MEs are kept busy.

In summary, if the power consumption is the dominant design factor, TDVS shall be a better choice. Otherwise, if performance is more important and packet loss needs to be avoided as much as possible, EDVS shall be used.

## 5 Conclusions

In this paper, we used an assertion-based design exploration methodology to study two different dynamic voltage scaling techniques in a network processor model: TDVS and EDVS. We analyzed the power and performance distributions with the two DVS policies and different parameter settings using automatically generated distribution analyzers based on assertion formulas. We studied the power-performance trade-offs with TDVS and EDVS applied and different thresholds and window sizes used. It was shown that in the NPU model the optimal configuration of a DVS policy usually depends on multiple factors such as the characteristics of the application, traffic loads and power or performance design requirements. The assertion-based analysis methodology was shown to be an efficient tool to help a designer choose an optimal configuration in a large design space, specially when the number of considered parameters is large and manual analysis of simulation results becomes tedious.

## References

[1] Y. Abarbanel, I. Beer, L. Gluhovsky, S. Keidar, and Y. Wolfsthal, “FoCs - automatic generation of simulation checkers from formal specifications”, *Technical Report, IBM Haifa Research Laboratory, Israel*, 2003.

[2] F. Balarin, Y. Watanabe, J. Burch, L. Lavagno, R. Passerone, and A. Sangiovanni-Vincentelli, “Constraints specification at higher levels of abstraction”, *International Workshop on High Level Design Validation and Test*, 2001.

[3] T. Burd and R. Brodersen, “Design issues for dynamic voltage scaling,” *International Symposium on Low Power Electronics and Design*, pp.9-14, 2000.

[4] X. Chen, H. Hsieh, F. Balarin, and Y. Watanabe, “Automatic trace analysis for logic of constraints”, *Design Automation Conference*, 2003.

[5] X. Chen, H. Hsieh, F. Balarin, and Y. Watanabe, “Verifying LOC Based Functional and Performance Constraints”,*International Workshop on High Level Design Validation and Test*, 2003.

[6] X. Chen, Y. Luo, H. Hsieh, L. Bhuyan, and F. Balarin, “Utilizing Formal Assertions for System Design of Network Processors”, *Design Automation and Test in Europe*, Feb. 2004.

[7] C. Eisner and D. Fisman. Sugar 2.0 proposal presented to the accellera formal verification technical committee. Mar. 2002.

[8] Y. Luo, J. Yang, L. Bhuyan, and L. Zhao, “NePSim: A Network Processor Simulator with Power Evaluation Framework”, *IEEE MICRO, special issue on network processors*, Sept., 2004.

[9] A. Pnueli, “The temporal logic of programs”, *The 18th IEEE Symposium on Foundation of Computer Science*, pages 46-57, 1977.

[10] http://developer.intel.com/design/network/ixa.html, Intel Corporation, IXP1200 Network Processor Family Hardware Reference Manual, 2001.

[11] http://developer.intel.com/design/intelxscale, Intel XScale microarchitecture, 2004.

[12] http://www.intel.com/design/network/products/npfamily/ixp2400.htm, Intel IXP2400 Network Processor, 2004.

[13] http://www.intel.com/design/network/products/npfamily/ixp2800.htm, Intel IXP2800 Network Processor, 2004.

[14] http://www.eda.org/vfv, PSL homepage, 2004.

[15] http://www.nlanr.net, the NLANR Measurement and Network Analysis, 2004.